\documentclass[conference,10pt]{IEEEtran}

\usepackage{graphicx}
\usepackage{subcaption}
\usepackage{caption}
\usepackage{booktabs}
\usepackage{tabularx}
\usepackage{multirow}
 \usepackage{nth}

\usepackage{graphicx}

\usepackage[backend=biber,style=ieee]{biblatex}
\AtBeginBibliography{\footnotesize}

\usepackage[binary-units=true]{siunitx}
\usepackage{glossaries}
\newacronym{pse}{PSE}{power sourcing equipment}
\newacronym{pd}{PD}{powered device}
\newacronym{usrp}{USRP}{universal software radio peripheral}
\newacronym{rpi}{RPi}{raspberry pi}
\newacronym{poe}{PoE}{power-over-Ethernet}
\newacronym{ptp}{PTP}{precision time protocol}
\newacronym{vlc}{VLC}{visible light communication}
\newacronym{vlp}{VLP}{visible light positioning}
\newacronym{tsn}{TSN}{time-sensitive networking}
\newacronym{sdr}{SDR}{software-defined radio}
\newacronym{los}{LoS}{line-of-sight}
\newacronym{pps}{PPS}{pulse per second}
\newacronym{gpclk}{GPCLK}{general purpose clock}
\newacronym{ros}{ROS}{robot operating system}
\newacronym{lco}{LCO}{lithium cobalt oxide}
\newacronym{tof}{ToF}{time-of-flight}
\newacronym{ntp}{NTP}{network time protocol}
\newacronym{soc}{SoC}{state of charge}
\newacronym{uhd}{UHD}{USRP hardware driver}
\newacronym{pcb}{PCB}{printed circuit board}
\newacronym{pll}{PLL}{phase-locked loop}
\newacronym{os}{OS}{operating system}
\newacronym{hat}{HAT}{hardware attached on top}
\newacronym{gpu}{GPU}{graphical processing unit}
\newacronym{cpu}{CPU}{central processing unit}
\newacronym{if}{IF}{intermediate frequency}
\newacronym{fpga}{FPGA}{field-programmable gate array}
\newacronym{rfic}{RFIC}{radio-frequency integrated circuit}
\newacronym{ssd}{SSD}{solid state drive}
\newacronym{vep}{VEP}{virtual edge platform}
\newacronym{wr}{WR}{white-rabbit}
\newacronym{gnss}{GNSS}{global navigation satellite system}
\newacronym{mems}{MEMS}{micro-electromechanical systems}
\newacronym{cf}{CF}{cell-free}
\newacronym{wpt}{WPT}{wireless-power transfer}
\newacronym{daq}{DAQ}{data acquisition system}
\newacronym{iot}{IoT}{Internet of Things}
\newacronym{dlc}{DLC}{Distributed Laser Charging}
\newacronym{ai}{AI}{Artificial Intelligence}
\newacronym{aoa}{AOA}{angle-of-arrival}
\newacronym{aod}{AOD}{angle-of-departure}
\newacronym{ap}{AP}{access point}
\newacronym{ar}{AR}{augmented reality}
\newacronym{asic}{ASIC}{application specific integrated circuit}
\newacronym{awgn}{AWGN}{additive white Gaussian noise}
\newacronym{ce}{CE}{channel estimation}
\newacronym{ced}{CED}{cumulative energy density}
\newacronym{csi}{CSI}{channel state information}
\newacronym{csp}{CSP}{contact service point}
\newacronym{crlb}{CRLB}{Cram\'er-Rao lower bound}
\newacronym{dc}{DC}{direct current}
\newacronym{dlt}{DLT}{Distributed Ledger Technology}
\newacronym{dp}{DP}{differencial privacy}
\newacronym{ecc}{ECC}{elliptic curve cryptography}
\newacronym{ecdlp}{ECDLP}{elliptic curve discrete logarithm problem}
\newacronym{ecsp}{ECSP}{edge computing service point}
\newacronym{eirp}{EIRP}{equivalent isotropically radiated power}
\newacronym{em}{EM}{electromagnetic}
\newacronym{en}{EN}{energy neutral}
\newacronym{e2e}{E2E}{End-to-End}
\newacronym{fa}{FA}{federation anchor}
\newacronym{fl}{FL}{federated learning}
\newacronym{gdpr}{GDPR}{general data protection regulation}
\newacronym{ic}{IC}{integrated circuit}
\newacronym{ioe}{IoE}{Internet of Everything}
\newacronym{id}{ID}{information decoding}
\newacronym{kpi}{KPI}{key performance indicator}
\newacronym{lca}{LCA}{life cycle assessment}
\newacronym{lis}{LIS}{large intelligent surface}
\newacronym{liion}{Li-ion}{lithium-ion}
\newacronym{lmo}{LMO}{lithium ion manganese oxide}
\newacronym{mf}{MF}{matched filter}
\newacronym{ml}{ML}{machine learning}
\newacronym{mmse}{MMSE}{minimum mean square error}
\newacronym{mmwave}{mmWave}{millimeter wave}
\newacronym{mimo}{MIMO}{multiple-input multiple-output}
\newacronym{miso}{MISO}{multiple-input single-output}
\newacronym{mpc}{MPC}{multipath component}
\newacronym{mrc}{MRC}{maximum ratio combining}
\newacronym{mrt}{MRT}{maximum ratio transmission}
\newacronym{ofdm}{OFDM}{orthogonal frequency-division multiplexing}
\newacronym{ota}{OTA}{over-the-air}
\newacronym{nb}{NB}{narrowband}
\newacronym{nr}{NR}{New Radio}
\newacronym{pa}{PA}{power amplifier}
\newacronym{pdp}{PDP}{power delay profile}
\newacronym{per}{PER}{packet error rate}
\newacronym{pet}{PET}{privacy enhancing technology}
\newacronym{pki}{PKI}{public key infrastructure}
\newacronym{pl}{PL}{path loss}
\newacronym{pc}{PC}{pilot count}
\newacronym{ran}{RAN}{radio access network}
\newacronym{rawc}{RAWC}{RIS-aided wireless communications}
\newacronym{rbit}{RBIT}{RIS-based information transmission}
\newacronym{re}{RE}{radio element}
\newacronym{rf}{RF}{radio frequency}
\newacronym{rfid}{RFID}{radio frequency identification}
\newacronym{ris}{RIS}{reflective intelligent surface}%reconfigurable?
\newacronym{rms}{RMS}{root mean square}
\newacronym{rss}{RSS}{received signal strength}
\newacronym{rssi}{RSSI}{received signal strength indicator}
\newacronym{rw}{RW}{RadioWeaves}
\newacronym{sa}{SA}{synchronization anchor}
\newacronym{sdg}{SDG}{Sustainable Development Goal}
\newacronym{sdn}{SDN}{software-defined network}
\newacronym{sinr}{SINR}{signal-to-interference-plus-noise ratio}
\newacronym{siso}{SISO}{single-input single-output}
\newacronym{slam}{SLAM}{simultaneous localization and mapping}
\newacronym{snr}{SNR}{signal-to-noise ratio}
\newacronym{svd}{SVD}{singular value decomposition}
\newacronym{tdd}{TDD}{time division duplexing}
\newacronym{tdoa}{TDOA}{time-difference-of-arrival}
\newacronym{toa}{TOA}{time-of-arrival}
\newacronym{ue}{UE}{user equipment}
\newacronym{uhf}{UHF}{ultra high frequency}
\newacronym{ula}{ULA}{uniform linear array}
\newacronym{upa}{UPA}{uniform planar array}
\newacronym{ura}{URA}{uniform rectangular array}
\newacronym{uv}{UV}{unmanned vehicle}
\newacronym{uwb}{UWB}{ultrawideband}
\newacronym{wb}{WB}{wideband}
\newacronym{zf}{ZF}{zero forcing}

\usepackage[capitalise]{cleveref}
\Crefformat{figure}{#2Fig.~#1#3}
\Crefmultiformat{figure}{Figs.~#2#1#3}{ and~#2#1#3}{, #2#1#3}{ and~#2#1#3}

\usepackage{xcolor}

\begin{document}
\title{Dynamic Federations for 6G Cell-Free Networking: Concepts and Terminology}

\author{
    \IEEEauthorblockN{
    Gilles Callebaut\IEEEauthorrefmark{1},  
    William T\"{a}rneberg\IEEEauthorrefmark{2},
    Liesbet Van der Perre\IEEEauthorrefmark{1},
    Emma Fitzgerald\IEEEauthorrefmark{2}
}
\IEEEauthorblockA{\IEEEauthorrefmark{1}
        KU Leuven, ESAT-Wavecore, Ghent Technology Campus, B-9000 Ghent, Belgium
       % gilles.callebaut@kuleuven.be
}
\IEEEauthorblockA{\IEEEauthorrefmark{2}
        Department of Electrical and Information Technology, Lund University, SE-221 00 Lund, Sweden
    }%
}

\IEEEspecialpapernotice{(Invited Paper)}

\maketitle

\begin{abstract}
Cell-Free networking is one of the prime candidates for 6G networks. Despite being capable of providing the 6G needs, practical limitations and considerations are often neglected in current research. In  this work, we introduce the concept of federations to dynamically scale and select the best set of resources, e.g., antennas, computing and data resources, to serve a given application.
Next to communication, 6G systems are expected to provide also wireless powering, positioning and sensing, further increasing the complexity of such systems. Therefore, each federation is self-managing and is distributed over the area in a cell-free manner. Next to the dynamic federations, new accompanying terminology is proposed to design cell-free systems taking into account practical limitations such as time synchronization and distributed processing. We conclude with an illustration with four federations, serving distinct applications, and introduce two new testbeds to study these architectures and concepts.
\end{abstract}


% \fi
\IEEEpeerreviewmaketitle

%\tableofcontents % overview of structure when writing

\section{Introduction}
The work on 6G has just begun. That effort includes addressing challenges in achieving very high data rates, imperceptibly low latency, unrivalled dependability, and ultra low power consumption \cite{REINDEERD1.1, Ericsson6g}. Importantly, the above should be realized while prioritizing the reduction of full networks' carbon footprint. To support 
the variety of 6G applications, wireless access architectures are proposed, hosting a high number of distributed radios and computing resources.
Cell-free networking~\cite{interdonato2019ubiquitous, NgoCellFree} provides an interesting concept to fully utilize the available capacity. While the theoretical potential of these novel architectures and networking paradigms have been recognized, many questions remain with respect to the feasibility of an actual deployment. 
%of the diversity and multitude of anticipated applications. 
How can such a great pool of resources get coordinated and allocated efficiently? How can both the infrastructure and the provided services be scalable? In this paper, we introduce the novel concept of \emph{dynamic federations}, that provides a key to addressing the above challenges. Dynamic federations consist of constellations of antennas, edge computing units, data storage, and other resources, to serve a specific application (class). %It establishes a dedicated logical architecture in the shared physical infrastructure. 
Moreover, we have identified the need to establish adequate terminology for the new distributed networking features, which we also introduce.

This paper is further organized as follows. In the next section, we explain how the 6G needs lead to the development of hyper-diverse connectivity platforms with distributed resources. In \Cref{sec:concepts} the new concept of dynamic federations is proposed, and consequently novel terminology is introduced in \Cref{sec:terminology}. An illustrative case study is elaborated in \Cref{sec:illustration}. Finally, \Cref{sec:conclusion} concludes this paper, inviting the R\&D community to discuss and adopt the novel concepts and terminology, and pointing out some plans for future work.

\begin{figure}[htbp]
    \centering
    \includegraphics[width=\linewidth]{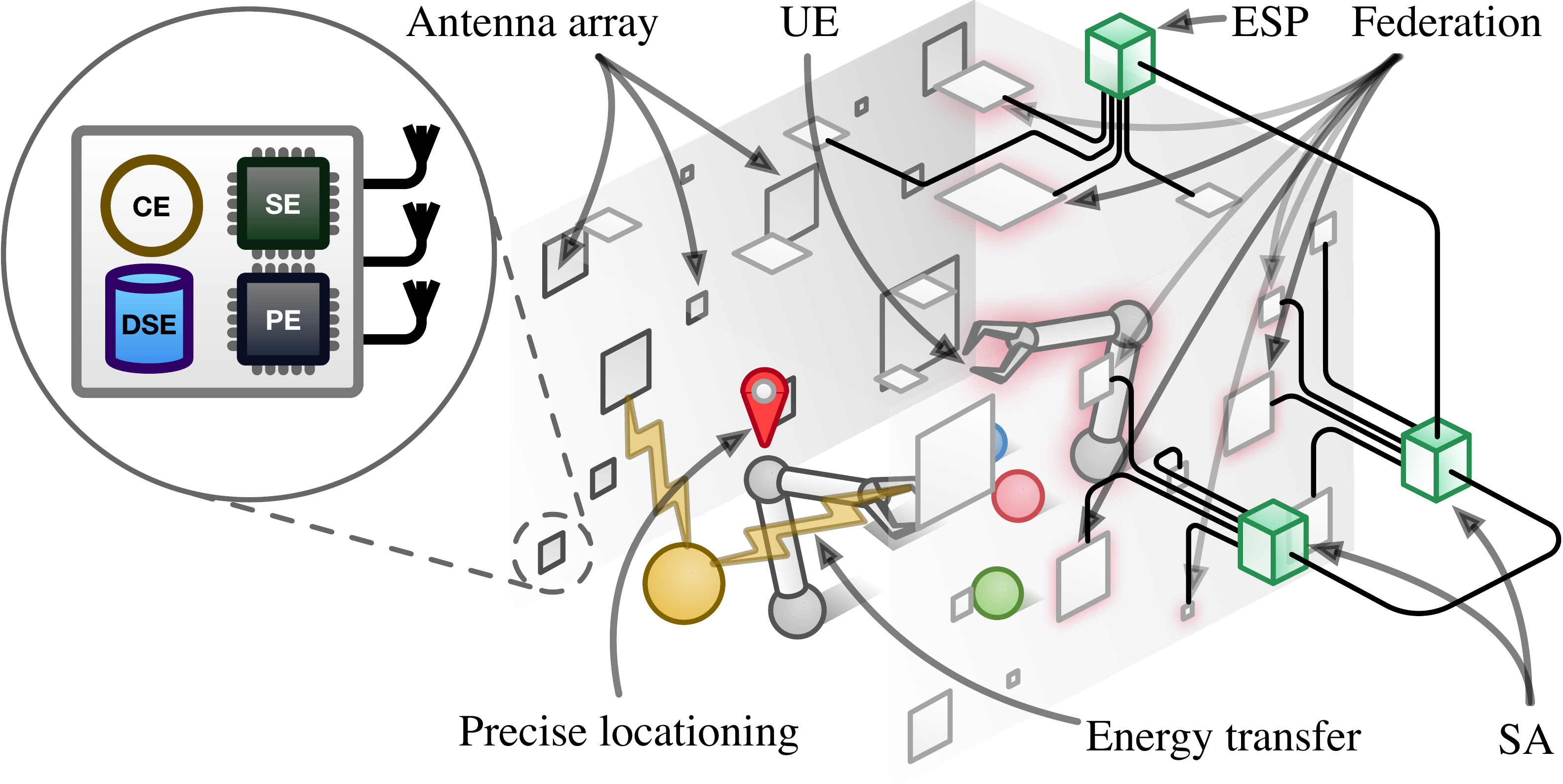}
    \caption{Example of a \acrlong{cf} system deployed in a Radioweaves setup~\cite{radioweaves}. The \gls{rw} system consists of at least one, but preferably many \glspl{ecsp}. This component is responsible for data aggregation and coordination of \glspl{csp}. A \gls{csp} is the first contact point from the \gls{ue} perspective and provides the necessary services to support user applications. A \gls{csp} can be equipped with one or several radio elements. To allow for a synchronized/coherent system, \gls{ecsp} can act as a synchronization anchor to synchronize its \glspl{csp}.
    % \gilles{depending on space, this should be reduced}\gilles{@William, could you make something like this?}\gilles{btw the combo of baseband processing and the antennes is the (E)CSP.}\william{Yes! I am on it. Just tying out the size.}
    }%
    \label{fig:structure}
\end{figure}

\section{6G Needs and Radio Access Architectures}\label{sec:platforms}

    \subsection{6G needs}\label{sec:6Gneeds}
        The \nth{6} generation of wireless networks will need to support a plethora of services. The number of connected devices and diversity of applications to be supported is expected to further increase. Notably, this diversity leads to new requirements for 6G infrastructures that have been analyzed in recent studies~\cite{Ericsson6g}. These include: 
        \begin{enumerate}
            \item The need to provide unprecedented capacity, better support for ultra reliability and imperceptible latency, and connecting a massive number of low power devices. This can be recognized as further raising the ambition along the three main axes considered in 5G.  
            \item Entirely new features to be supported by the network infrastructure. In particular, (i) many applications will rely on position information and (ii) connections to energy-neutral devices will need to be established. 
        \end{enumerate}
        In particular, novel resilient applications will rely on 'real-time' and 'real-space' interaction, whereby the physical and virtual realities share the same temporal and spatial reference frame. This necessitates a new type of radio access architecture and a paradigm shift in the wireless networking. Dense infrastructures hosting a very large number of distributed radios and computational resources bear a great potential. Such systems can offer ample diversity, solid redundancy, and proximity that can increase the link energy efficiency with orders of magnitude~\cite{Ganesan}.  Distributed computation and storage can also provide nearby data processing and reduction. Many have introduced technologies that can provide the distributed infrastructure and resources, as discussed below.

    % \liesbet{Reorganizing/Clarifying the problems}
    
    \subsection{Next-Generation Radio Access Architectures}
        The envisioned radio access systems will need to support extreme data rates, imperceptibly low latency, dependability on par with wired networks, low-power usage, wireless energy transfer for energy neutral devices, and precise positioning~\cite{Ericsson6g}. 
        
        Several radio access concepts to address these new requirements are proposed in literature. %These architectures are designed to combat, e.g., the unfavorable wireless propagation environment. To do this
        For instance, \gls{ris}, where a large two-dimensional array consisting of large number of reflective elements is used to reflect signals to a desired location. This effectively allows controlling the wireless environment. Two implementations of \gls{ris} are proposed: \gls{rawc}~\cite{9551980} and \gls{rbit}~\cite{lin2020reconfigurable}. The former uses passive reflectors to do interference suppression and signal steering, while the latter acts as a backscatter where the signals are modulated by the \gls{ris} array. Another approach is to further extend the concept of distributed massive MIMO to the \acrlong{cf} case. In \gls{cf} massive MIMO, a set of geographically distributed antennas jointly serve a number of users. There are no cell boundaries because all \glspl{ap}, equipped with one or multiple antennas, are connected to a \gls{cpu} coordinating the system. This approach is impractical to deploy in real scenarios. A special case of a \acrlong{cf} system is \gls{lis}~\cite{hu2018beyond}, where a large set of active antennas are densely distributed throughout a three-dimensional space. Where in \gls{cf} the antennas are geographically distributed, in \gls{lis} large panels of resources are encapsulating the users. The antennas are managed in a distributed and cell-free manner, i.e., all resources in the network can be used to provide a given service. Ongoing \gls{lis} research~\cite{radioweaves} is aimed at geographically constrained spaces, e.g., sports arenas or factories. However, the concept itself allows for much larger networks where users can seamlessly move around in the infrastructure, such as in a city, without requiring the hand-overs as in non-cell-free systems.

        The systems described above all focus on wireless communication, however, 6G will also support, among others, precise positioning services~\cite{hu2018beyond} and wireless energy transfer~\cite{lopez2021massive}, supporting energy-neutral \gls{iot} devices~\cite{Deut2205:Location}. This is the emphasis of this work. A system architecture, having these new services in mind, is \acrlong{rw}~\cite{REINDEER,callebaut2022techtile} and is further elaborated in the following section.
        
        % Many antennas, distributed over a wide area but controlled together, give a large aperture for positioning that can achieve far greater accuracy than existing antenna array-based positioning systems. For wireless power transfer, the closer the antennas are to the users, the better, since even in free space the transferred power drops off quickly with distance, and this is exacerbated by multipath effects. With \gls{lis}, a user is likely to be close to several antenna panels at any given time, even for mobile users, yielding an improvement in efficiency of the wireless power transfer.

        % In envisioned \gls{lis} systems, such as \gls{rw}~\cite{REINDEER,callebaut2022techtile}, groups of distributed antennas are aggregated by \emph{baseband} units. The baseband units carry out signal processing, such as beam-forming, and are often implemented in ASIC or \glspl{fpga}. These units are interconnected at a point where the system can be managed and traffic routed, and from which the system can be connected to the wider Internet.

\section{Dynamic federations: concepts for 6G cell-free networking}\label{sec:terminology}

%\subsection{Dynamic Federation Concepts}

%\section{Dynamic Federations: Terminology}\label{sec:terminology}

    \subsection{Concepts}\label{sec:concepts}
        As discussed in Section~\ref{sec:6Gneeds}, a further evolution of current network architectures will not be able to support the anticipated requirements for 6G, therefore necessitating novel terminology and concepts. The currently studied \gls{cf} systems will need to be constrained in order to allow practical deployments. For example, the notion of one central processing unit and not having cells will become impractical. %Consequently, to design adequate practical \gls{cf} and \gls{rw} systems, we introduce new terminology and concepts.  
        
        Due to the rich set of diverse services, the set of resources used, e.g., processing and radio elements, are tailored to the particular application. This means that the wireless access infrastructure needs to allocate resources to specific applications with different requirements. For instance, for wireless power transfer, charging resources located close to the intended device will yield the highest efficiency. In contrast, in XR applications the mobility of the user's body in space and the head movement, requires a high spatial diversity of the antenna resources to mitigate outage and peaks in latency. This demonstrates the need for grouping of resources in a \acrlong{cf} context in a dynamic manner, i.e., in both the temporal and spatial domain. In this work, we introduce the term \emph{federation(s)} to denote the group of resources which jointly serve a given application. In the next part, the terminology is introduced, which is required in order to design these \gls{rw} systems taking practical implications and 6G needs into account.

\subsection{Terminology}\label{sec:terminology}
Next-generation networks are required to shift from supporting only communication to also provide sensing, positioning and wireless power transfer~\cite{callebaut2022techtile}. This entails that several additional resources need to be embedded in the core functionality of the architecture. This, next to the concept of dynamic federations, requires a specific set of terminology in order to devise such systems in practice. In this paper, we distinguish logical entities and physical elements. The latter denotes hardware elements present in the infrastructure to support, e.g., wireless charging. These hardware resources can be logically mapped to entities to form the \gls{rw} infrastructure and contact points for the users. An example of such an implementation is shown in \cref{fig:structure}. The physical and logical structure, including its hierarchical components, are depicted in \cref{fig:architecture}. The newly introduced terms are described as generally as possible to not impose any constraints on the implementation of these systems.

% An overview of the main components in such an infrastructure is shown in Figure~\ref{fig:structure}. An explicit distinction is made between logical and physical components, which we propose to call logical entities and physical elements, respectively. The physical and logical structure, including its hierarchical components, are depicted in Figure~\ref{fig:architecture}. The newly introduced terms are described as general as possible to not impose any constraints on the implementation of these systems. 

% \begin{itemize}
%     \item Practical systems are constrained in the connections between different geographically separated antennas or antenna arrays. 
%     \item new services are added
%     \item no clear separation between front-haul and back-haul
%     \item embedded Local connectivity-computational resources 
%     \item orchestration of local clusters or groups, i.e., federations, are required to balance processing and data aggregating load.
%     \item we propose a new set of terminology to accommodate these novel concepts.
% \end{itemize}

\subsubsection{Logical Entities}
The network consists of several logical components, elaborated in Table~\ref{tab:terms}. The first entity seen from the perspective of the \gls{ue} is the \textit{\acrlong{csp}}. This logical service point allows to power the device wired or wirelessly, provide wireless communication or could host other elements. Several \glspl{csp} are connected to one or more \glspl{ecsp}. The \gls{ecsp} can have a dedicated connection to the back-haul and other \glspl{ecsp}. The task of the \gls{ecsp} is to provide dedicated computing resources used for collective tasks such as, e.g., coherent channel-matched beamforming. Also, this entity can host several hardware elements besides processing/memory. 

As not all \glspl{csp} will contribute equally to all services, the notion of a federation is introduced. During the operation, federations will be orchestrated depending on the served \glspl{ue} and their application classes, the propagation environment, and the load on the \glspl{csp}. A federation is a collection of \glspl{csp} jointly serving one or multiple \glspl{ue}. A federation is typically coordinated by an \gls{ecsp} acting as the \acrlong{fa}. Often, federations will consist of \glspl{csp} located closely together, but this is not mandated, nor desired in some cases.

\begin{table*}[]
    \centering
    \caption{Logical Entities and Physical Elements}\label{tab:terms}
    \begin{tabularx}{\textwidth}{c l l X}
    \toprule
    & Name & Abbreviation & Description\\
    \midrule\\
    \multirow{22}{*}{\rotatebox[origin=c]{90}{Logical Entities}}
        &\Acrlong{rw} & \acrshort{rw} & Wireless access infrastructure consisting of a fabric of distributed radio, computing, and storage resources.\\
         &\Acrlong{csp} & \acrshort{csp} & Integrates local computation and storage resources, and provides at least communication, sensing or charging functionality. It is the first contact point as seen from the \gls{ue} and takes the role of an anchor in the context of position related applications. \\
         &\Acrlong{ecsp} & \acrshort{ecsp} & Shared compute resources integrated in the \gls{rw} that can support applications in need of substantial compute power and/or connection to the back-haul or other \gls{rw} infrastructures.\\
         & Federation & - & (Temporary) set of cooperating resources in the \gls{rw}, working in unison, that could be more or less synchronized, and including at least \glspl{csp} and typically a synchronization anchor, and potentially edge processing unit(s), established to serve a cluster of devices and/or application(s).\\
         &\Acrlong{sa} & \acrshort{sa} &Logical function flexibly located attributed to a certain \gls{csp} to serve as a synchronization reference for a set of cooperating \glspl{csp} for some period.\\
         &\Acrlong{fa} & \acrshort{fa} & The \gls{fa} is responsible to orchestrate and to coordinate a federation. This task will be primarily performed by an \gls{ecsp}.\\
         &\Acrlong{en} device(s) & \acrshort{en}-device(s) & \Gls{en} devices are a specific subset of \glspl{ue}, housing dedicated circuitry for energy harvesting. Their main characteristic is that they are passive devices, i.e., they do not have their own power supply.
    All power they use for operation is harvested from incident \gls{em} fields.
    From a perspective of \gls{em} fields, they act as a power sink, 
    as opposed to devices that have some internal power supply.
    \gls{en} devices rely on the \gls{wpt} capabilities of the infrastructure for power provisioning. 
    In contrast to conventional networks, \gls{rw} inherently supports \gls{en} devices, requiring dedicated protocols and technologies to do so. 
    This includes both energy harvesting techniques with intentional sources (i.e., \gls{wpt}) and with unintentional sources (i.e., ambient energy harvesting)\\
    
     \\ 
    \midrule
    \\ 
    
    \multirow{12}{*}{\rotatebox[origin=c]{90}{Physical Elements}}
    & Sensing Element & & Unit integrated in a \gls{csp} that can sense signals in the environment via radio channels or other media, e.g., a camera.\\
    
    & Data Storage Element &&  Memory resource integrated in a \gls{csp}.\\
    
    &  Processing Element &&Local computational resources integrated in a \gls{csp}.\\
    
    & Charging Element &&Functionality integrated in a \gls{csp} that can efficiently charge devices in the environment, e.g., electromagnetic via antennas or inductive through coils.\\
    
    & Radio Element &&Transmit/receive units, most often including an antenna, that can serve to exchange data or charge devices using electromagnetic waves. \\
    
    & X-haul && It interconnects \glspl{csp} locally (front-haul) and also provides access to remote network and cloud resources (back-haul). It can comprise both wired (including optical fibers connections) and wireless segments. In contrast to conventional networks, in \gls{rw}, no clear distinction can be made between the front- and back-haul. The X-haul, thus, comprises a mix of the two.\\
    
         \bottomrule
    \end{tabularx}
\end{table*}

\subsubsection{Physical Elements}
\begin{figure}[hbtp]
    % \hfill
    % \begin{subfigure}[t]{0.48\linewidth}
    % \centering
    %     \includegraphics[height=0.55\linewidth]{figures/logical-arch.png}
    % \end{subfigure}\hfill
    % \begin{subfigure}[t]{0.48\linewidth}
    % \centering
    %     \includegraphics[height=0.55\linewidth]{figures/physical-arch.png}
    % \end{subfigure}
    % \hfill
    \includegraphics[width=\linewidth]{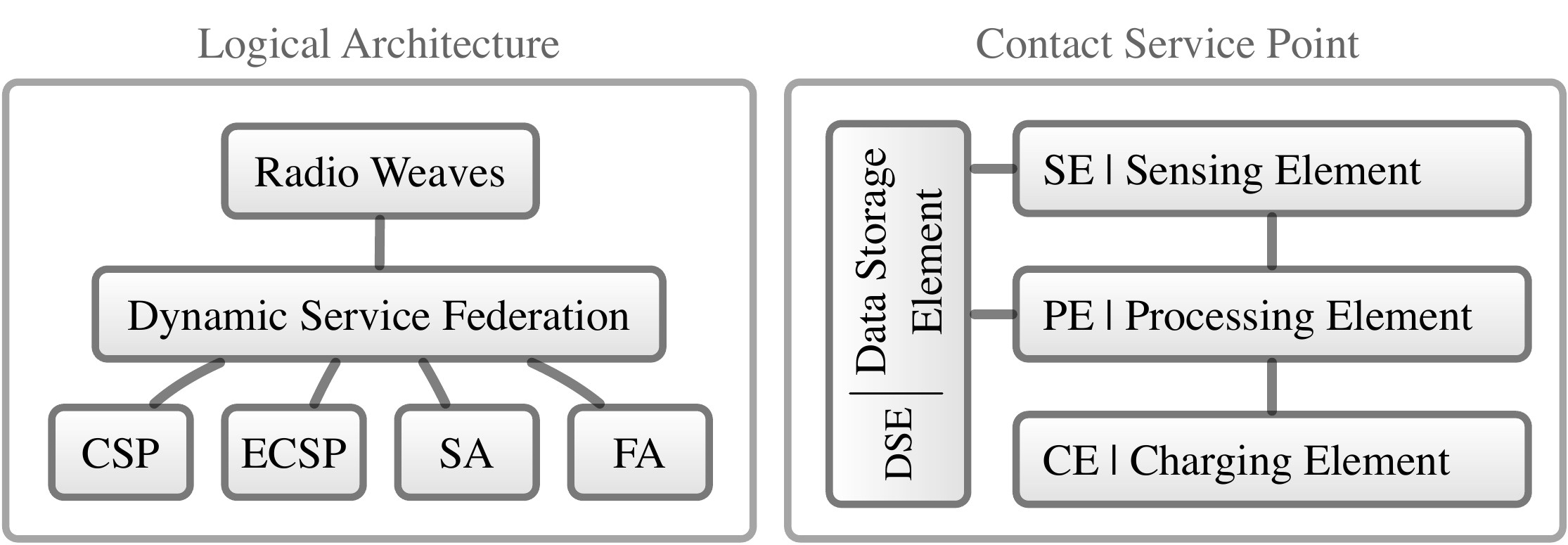}
    \caption{Physical and logical architecture of the \gls{rw} setup. The physical architecture depicts an example implementation of a \gls{csp} including data storage, sensing, processing, charging and radio elements.}
    %\gilles{could we also Williamfy this? to have a consistent set of figures}} \william{On it.}
    \label{fig:architecture}
\end{figure}

The logical components consist of several physical elements.
As in a \gls{rw} infrastructure the network provides not only communication but also positioning and power transfer, the service points consist of more than only radio elements. All physical components are summarized in Table~\ref{tab:terms}. An example of the (optional) physical components hosted on a \gls{csp} is shown in~\cref{fig:architecture}.

    % \textbf{Sensing Element} Unit integrated in a \gls{csp} that can sense signals in the environment via radio channels or other media, e.g., a camera.
    
    % \textbf{Data Storage Element} Memory resource integrated in a \gls{csp}.
    
    % \textbf{Processing Element} Local computational resources integrated in a \gls{csp}.
    
    % \textbf{Charging Element} Functionality integrated in a \gls{csp} that can efficiently charge devices in the environment, e.g., electromagnetic via antennas or inductive through coils.
    
    % \textbf{Radio Element} Transmit/receive units, most often including an antenna, that can serve to exchange data or charge devices using electromagnetic waves. 
    
    % \textbf{X-haul} It interconnects \glspl{csp} locally (front-haul) and also provides access to remote network and cloud resources (back-haul). It can comprise both wired (including optical fibers connections) and wireless segments. In contrast to conventional networks, in \gls{rw}, no clear distinction can be made between the front- and back-haul. The X-haul, thus, comprises a mix of the two.

\section{Illustration}\label{sec:illustration}

\begin{figure}
    \centering
    \includegraphics[width=\columnwidth]{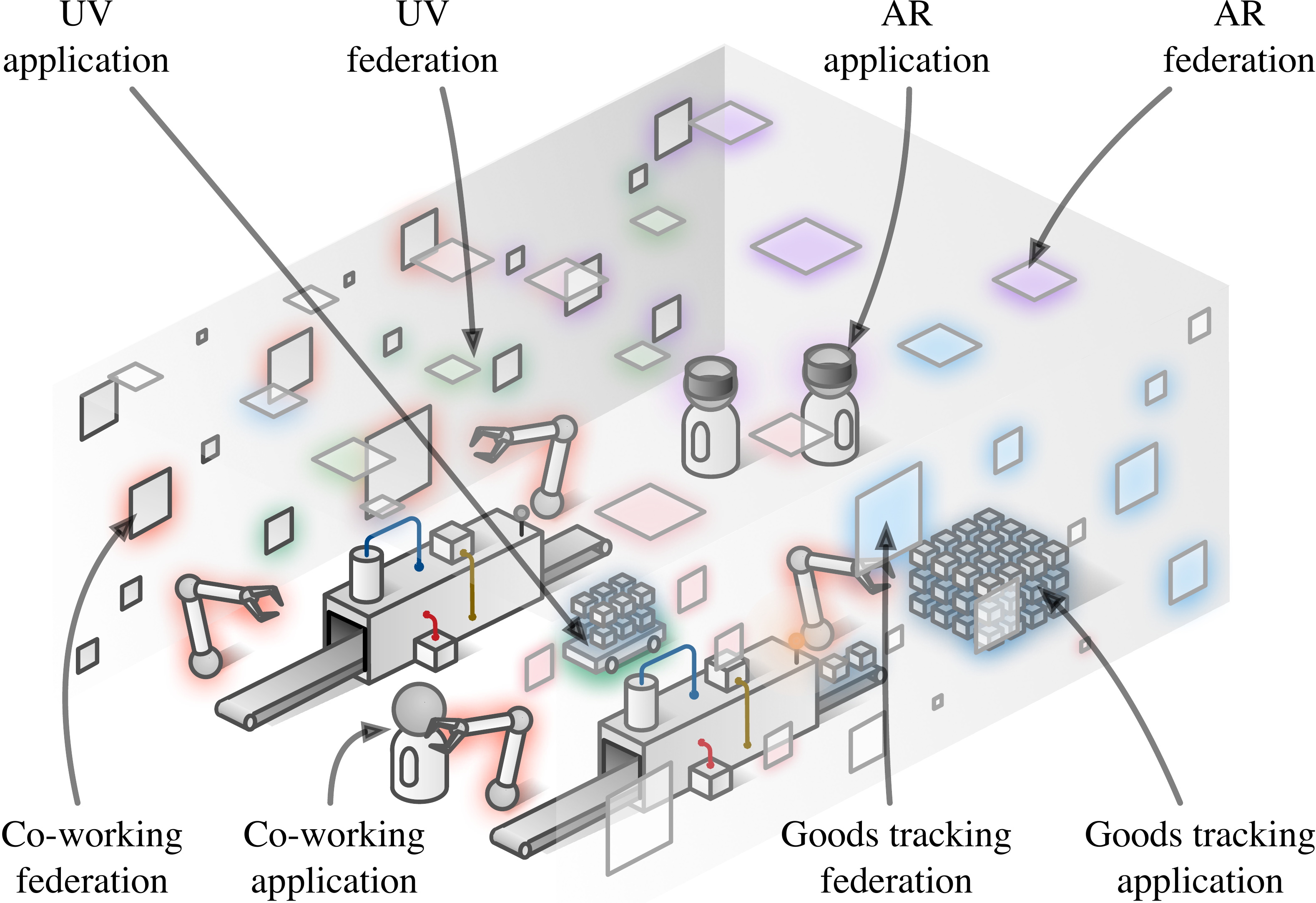}
    \caption{An example \gls{rw} deployment in a smart factory, with federations and their served devices color coded. The four applications are \gls{ar} for professional applications (purple), tracking of robots and \glspl{uv} (green), tracking of goods and real-time inventory (blue), and human-robot co-working (red). %\gilles{no humans working with the red robots}
    % \william{I think this figure is cluttered, a curved space/wall would be cognitively more comprehensible, but very time-consuming to draw. I'll see if I get some spare hours.}\gilles{i think it is ok. It also illustrates, by being cluttered, that a lot needs to happen an the federations are not condensed in specific "hot spots", but could be rather sparse.}
    % \william{Needs labels. But not too many.}\gilles{maybe just adapt mine with labels? and keep this one as is?} \william{Keep both? I don't quite follow.}\gilles{Prev. remark was before the other fig was removed. Keeping this as is, is fine now :)}
    }\label{fig:federations_example}
\end{figure}

\Cref{fig:federations_example} shows an example deployment of RadioWeaves in a smart factory, with four federations, shown in different colors. RadioWeaves \glspl{csp} are deployed throughout the production hall on the walls and ceiling, and are dynamically assigned to federations to serve the devices and their running applications that are present at any given time. The constellation of \glspl{csp} assigned to each federation is tailored to the particular application's requirements.

% Placeholder, to be replaced with a proper digital figure.

In \Cref{fig:federations_example}, there are currently four different applications running, taken from the use cases presented in \cite{REINDEERD1.1}. These are \gls{ar} for professional applications (shown in purple), tracking of robots and \glspl{uv} (green), tracking of goods and real-time inventory (blue), and human-robot co-working (red).
Each application is served by a federation, with its \glspl{csp} shown in the same color as the application.

For the \gls{ar} for professional applications use case, human workers wear energy-neutral \gls{ar} goggles, which display digital information overlaid on the physical area in which they are working. As described in~\cite{REINDEERD1.1}, using energy-neutral devices allows the goggles to be extremely light and thus comfortable to wear, but it increases the requirements on the infrastructure, which must transmit uncompressed video to the goggles as they lack the processing capability to decode a compressed video stream. This means that this application requires an extremely high data rate (potentially up to \SI{3}{Gbps}~\cite{REINDEERD1.1}), as well as very low latency to prevent motion sickness. The goggle also need to be powered via wireless power transfer from the serving federation. To meet these requirements, the federation thus consists of a cluster of \glspl{csp} located on the wall and ceiling close to the user devices. The close proximity of the \glspl{csp} gives a good link budget both for communication and wireless power transfer, and the short distances between \glspl{csp} keeps the latency low.

Meanwhile, for the UV tracking use case, one of the biggest challenges is mobility. Tracking of the \gls{uv} also requires low latency, high reliability, and a relatively high data rate. To serve this application, we again show a cluster of nearby \glspl{csp}, but in order to account for the \glspl{uv} mobility, the federation is adapted as the \gls{uv} moves. Some \glspl{csp} are not currently serving the \gls{uv} but are standing by to join the federation as it is predicted the \gls{uv} will move in their direction. This dynamic adaptation of the federation as it ``follows'' the robot around the factory floor ensures a consistently good channel even as the robot passes objects that may cause shadowing, while also allowing the federation's panels to be physically close to each other to provide low latency communications.

In the tracking of goods use case, the requirements on latency, data rate, and reliability are significantly relaxed, but instead positioning accuracy becomes the most important requirement as the items to be tracked move around the production hall. A key ingredient in providing high-accuracy positioning is aperture size, and so for this application the federation is assigned panels spread out over the deployment area. The spatial diversity thus provided also ensures goods can be located and the tracking devices communicated with anywhere in the production hall.

Finally, for the human-robot co-working use case, we again see a dense cluster of \glspl{csp} located close to the user devices. As in the AR use case, this provided good data rates, high reliability, and low latency. This use case also does not require wireless power transfer, as it does not employ energy-neutral devices. This simplifies initial access for the devices because the devices themselves can actively contact the infrastructure, rather than needing to receive sufficient power to establish initial contact before their position can be determined. This, combined with a somewhat lower data rate compared with the AR use case, means that fewer \glspl{csp} are needed per user for the human-robot co-working federation.

\section{Conclusion and future work} \label{sec:conclusion}

To support the challenging 6G use cases that are emerging, a distributed, cell-free architecture will be needed, if not universally then at least in specific application domains with dense deployments. We propose the introduction of dynamic federations, consisting of constellations of antennas, edge computing units, data storage, and other resources, to serve specific applications or application classes. Each federation can be distributed throughout the deployment area in a cell-free manner, and internally manages itself, simplifying orchestration in such a complex system. We have developed an accompanying terminology, with definitions for physical and logical entities, to describe this type of architecture and its federations. We invite the research and development community to discuss, further develop, and adopt both the federation concept and the terminology for distributed, cell-free 6G systems.

We plan to bring such an architecture, along with dynamic federation orchestration, into reality by implementing them in two testbeds, located at KU Leuven and Lund University. The KU Leuven testbed, Techtile~\cite{callebaut2022techtile}, was inaugurated in October 2021 and already has the physical infrastructure in place, but work is ongoing on the software to run the testbed. Techtile consists of a room built with 140 modular panels, each of which contains software-defined radios, edge computing units, and sensors. The Lund University testbed is scheduled to become operational in mid-2023 and will consist of a number of Xilinx UltraScale Plus radio frequency system-on-chip devices, each of which contain 16 antenna elements, along with X-haul connectivity between them and co-located edge computing resources. With these two testbeds, we will be able to implement the federation concept and test its performance in real application scenarios.


\section*{Acknowledgment}
The project has received funding from the European Union's Horizon 2020 research and innovation programme under grant agreement No 101013425.

The authors would like to thank the REINDEER team for the rich discussions that have strengthened the definition of the new terminology.

{\footnotesize \printbibliography}

@inproceedings{radioweaves,
	title        = {{RadioWeaves for efficient connectivity: analysis and impact of constraints in actual deployments}},
	author       = {Van der Perre, Liesbet and Larsson, Erik G and Tufvesson, Fredrik and De Strycker, Lieven and Bjornson, Emil and Edfors, Ove},
	year         = 2019,
	journal      = {CONFERENCE RECORD OF THE 2019 FIFTY-THIRD ASILOMAR CONFERENCE ON SIGNALS, SYSTEMS & COMPUTERS},
	publisher    = {IEEE},
	pages        = {15--22},
	issn         = {1058-6393},
	language     = {eng},
	organization = {Matthews, MB}
}

@online{REINDEERD1.1,
	title        = {{REsilient INteractive applications through hyper Diversity in Energy Efficient RadioWeaves technology (REINDEER) project -  Deliverable 1.1: Use case-driven specifications and technical requirements and initial channel model}},
	author       = {{EU H2020 REINDEER project}},
	year         = 2021,
	url          = {https://reindeer-project.eu/D1.1},
	urldate      = 2021,
	note         = {Visited on 2021-07-26}
}

@inproceedings{Ganesan,
	title        = {{RadioWeaves for Extreme Spatial Multiplexing in Indoor Environments}},
	author       = {Ganesan, Unnikrishnan Kunnath and Björnson, Emil and Larsson, Erik G.},
	year         = 2020,
	booktitle    = {2020 54th Asilomar Conference on Signals, Systems, and Computers},
	volume       = {},
	number       = {},
	pages        = {1007--1011},
	doi          = {10.1109/IEEECONF51394.2020.9443342}
}

@article{NgoCellFree,
	title        = {{Cell-Free Massive MIMO Versus Small Cells}},
	author       = {Ngo, Hien Quoc and Ashikhmin, Alexei and Yang, Hong and Larsson, Erik G. and Marzetta, Thomas L.},
	year         = 2017,
	journal      = {IEEE Transactions on Wireless Communications},
	volume       = 16,
	number       = 3,
	pages        = {1834--1850},
	doi          = {10.1109/TWC.2017.2655515}
}

@misc{Ericsson6g,
	title        = {{Joint communication and sensing in 6G networks}},
	author       = {{Ericsson AB}},
	note         = 2021,
	howpublished = {https://www.ericsson.com/en/6g}
}

@INPROCEEDINGS{Deut2205:Location,
AUTHOR="Benjamin J. B. Deutschmann and Thomas Wilding and Erik G. Larsson and Klaus
Witrisal",
TITLE="Location-based Initial Access for Wireless Power Transfer with Physically
Large Arrays",
BOOKTITLE="WS08 IEEE ICC 2022 Workshop on Synergies of communication, localization,
and sensing towards 6G (WS08 ICC'22 Workshop - ComLS-6G)",
ADDRESS="Seoul, Korea (South)",
MONTH=may,
YEAR=2022
}

@article{hu2018beyond,
	title        = {{Beyond massive MIMO: The potential of positioning with large intelligent surfaces}},
	author       = {Hu, Sha and Rusek, Fredrik and Edfors, Ove},
	year         = 2018,
	journal      = {IEEE Transactions on Signal Processing},
	publisher    = {IEEE},
	*volume      = 66,
	**number     = 7,
	*pages       = {1761--1774}
}

@misc{callebaut2022techtile,
	title        = {{Techtile -- Open 6G R\&D Testbed for Communication, Positioning, Sensing, WPT and Federated Learning}},
	author       = {Gilles Callebaut and Jarne Van Mulders and Geoffrey Ottoy and Daan Delabie and Bert Cox and Nobby Stevens and Liesbet Van der Perre},
	year         = 2022,
	eprint       = {2202.04524},
	archiveprefix = {arXiv},
	primaryclass = {eess.SP}
}

@article{lin2020reconfigurable,
	title        = {{Reconfigurable intelligent surfaces with reflection pattern modulation: Beamforming design and performance analysis}},
	author       = {Lin, Shaoe and Zheng, Beixiong and Alexandropoulos, George C and Wen, Miaowen and Di Renzo, Marco and Chen, Fangjiong},
	year         = 2020,
	journal      = {IEEE Transactions on Wireless Communications},
	publisher    = {IEEE},
	volume       = 20,
	number       = 2,
	pages        = {741--754}
}

@article{interdonato2019ubiquitous,
	title        = {{Ubiquitous cell-free massive MIMO communications}},
	author       = {Interdonato, Giovanni and Bj{\"o}rnson, Emil and Ngo, Hien Quoc and Frenger, P{\aa}l and Larsson, Erik G},
	year         = 2019,
	journal      = {EURASIP Journal on Wireless Communications and Networking},
	publisher    = {Springer},
	*number      = 1,
	*pages       = {1--13},
	*volume      = 2019,
	date-added   = {2021-10-04 15:05:22 +0200},
	date-modified = {2021-10-04 15:05:22 +0200}
}

@misc{REINDEER,
	title        = {{REINDEER (REsilient INteractive applications through hyper Diversity in Energy Efficient RadioWeaves technology)}},
	note         = {Online, accessed 2021-07-30},
	howpublished = {https://reindeer-project.eu/}
}

@misc{LIS,
	title        = {{LIS (Large Intelligent Surfaces -- Architecture and Hardware)}},
	note         = {Online, accessed 2021-07-30},
	howpublished = {https://strategiska.se/en/research/ongoing-research/ssf-computing-and-hardware-infrastructure-2019/project/10873/}
}

@ARTICLE{9551980,
  author={Pei, Xilong and Yin, Haifan and Tan, Li and Cao, Lin and Li, Zhanpeng and Wang, Kai and Zhang, Kun and Björnson, Emil},
  journal={IEEE Transactions on Communications}, 
  title={RIS-Aided Wireless Communications: Prototyping, Adaptive Beamforming, and Indoor/Outdoor Field Trials}, 
  year={2021},
  volume={69},
  number={12},
  pages={8627-8640},
  doi={10.1109/TCOMM.2021.3116151}}

@article{lopez2021massive,
	title        = {{Massive Wireless Energy Transfer: Enabling Sustainable IoT Toward 6G Era}},
	author       = {L{\'o}pez, Onel LA and Alves, Hirley and Souza, Richard Demo and Montejo-S{\'a}nchez, Samuel and Fern{\'a}ndez, Evelio Martin Garcia and Latva-Aho, Matti},
	year         = 2021,
	journal      = {IEEE Internet of Things Journal},
	publisher    = {IEEE}
}

\end{document}